# Cyber-Cosmos: A New Citizen Science Concept in A Dark Sky Destination

**Domingos Barbosa[a]\*, Bruno Coelho[a]\*, Miguel Bergano[a], Catarina Magalhães[c], David Mendonça[d], Daniela Silva[c], Alexandre C. M. Correia[e], João Pandeirada[a,b] ,Valério Ribeiro[a], Thomas Esposito[f,g,h], Franck Marchis[f,g]**

[a] *Instituto de Telecomunicações, Campus Universitário de Santiago, 3810-193 Aveiro, Portugal;*
dbarbosa@av.it.pt*(DB);* brunodfcoelho@av.it.pt*(BC);* jbergano@av.it.pt*(MB);* joao.pandeirada@ua.pt*(JP);* valerio.alipio.ribeiro@gmail.com*(VR);*
[b] *Department of Electronics, Telecommunications and Informatics, University of Aveiro, 3810-193 Aveiro, Portugal*
[c] *Agrupamento de Escolas da Escalada, Bairro de S. Martinho, 3320-206 Pampilhosa da Serra, Portugal;* daniela.ssilva55@gmail.com(DS);catarinapixi5@gmail.com(CM)
[d] *Agrupamento de Escolas Amato Lusitano, Av. Pedro Alvares Cabral 10, 6000-085 Castelo Branco, Portugal;* davidgrau2126@gmail.com *(DM)*
[e] *CFisUC, Departamento de Física, Universidade de Coimbra, 3004-516 Coimbra, Portugal;* acor@uc.pt *(AC)*
[f] *SETI Institute, 189 Bernardo Ave, Suite 200, Mountain View, CA 94043, United States of America;* fmarchis@seti.org *(FM);* tesposito@seti.org *(TE)*
[g] *Unistellar SAS, 19 Rue Vacon, 13001 Marseille, France*
[h] *Department of Astronomy, 501 Campbell Hall MC #3411, University of California, Berkeley, CA 94720, United States of America*
\* Corresponding Authors

**Abstract**

Astrotourism and related citizen science activities are becoming a major trend of a sustainable, high-quality tourism segment, core elements to the protection of Dark skies in many countries. In the Summer of 2020, in the middle of COVID pandemics, we started an initiative to train young students - Cyber-Cosmos - using an Unistellar eVscope, a smart, compact and user-friendly digital telescope that offers unprecedented opportunities for deep-sky observation and citizen science campaigns. Sponsored by the Ciência Viva Summer program, this was probably the first continuous application of this equipment in a pedagogical and citizen-science context, and in a pandemic context. Pampilhosa da Serra, home to a certified Dark Sky destination (Aldeias do Xisto) in central Portugal, was the chosen location for this project, where we expect astrotourism and citizen science to flourish and contribute to space sciences education.

**Keywords:** astrotourism; citizen science; education; project-based learning; STEAM; dark sky; planetary sciences, space situational awareness

## 1. Introduction

Astrotourism is a major trend of a sustainable, high-quality tourism segment, a recognized central element in the protection of Dark skies in many countries. In Portugal, several Dark Sky destinations have appeared in recent years following a growing awareness of its potential to boost the local economy by regional stakeholders around professional observatories or simply to benefit stargazing activities as part of ecotourism activities. It benefits from the dark skies of those sites and requires little infrastructure to get stargazing started. It is also backed by a growing desire from the general public to experience rare night skies. The International Dark-Sky Association (IDA) has championed the protection of the night skies for present and future generations with pioneering activities against light pollution [1]. IDA promotes win-win solutions that allow people to appreciate dark, star-filled skies while enjoying the benefits of responsible outdoor lighting. Stressing these beneficial development aspects, UNESCO, the United Nations World Tourism Organization (UNWTO) and the International Astronomical Union (IAU) have jointly recognized the benefits of protection of Dark Skies and have developed guidelines and promoted certifications like the Dark Sky Starlight Foundation Tourism Destinations to protect and defend the sky and valuing it as a resource necessary for life and the intangible heritage of humanity [2,3]. The "La Palma Declaration" emphasises the protection and conservation of the night skies as an important scientific, cultural, environmental and tourist resource [2]. Therefore, the Dark Sky territories or destinations are prime locations to initiate citizen science projects and connect astrotourism initiatives to science education. The benefits of these initiatives have a recognized social and economic impact that contribute to the development as outlined through the United Nations' Sustainable Development Goals (SDGs).

As described in [4], the SDGs impacted by astrotourism and related science education activities include at least SDGs 4, 5, 8, 9, 10, and 11: quality education; gender equality; decent work and economic growth; industry, innovation, and infrastructure; reduced inequalities; and sustainable communities.

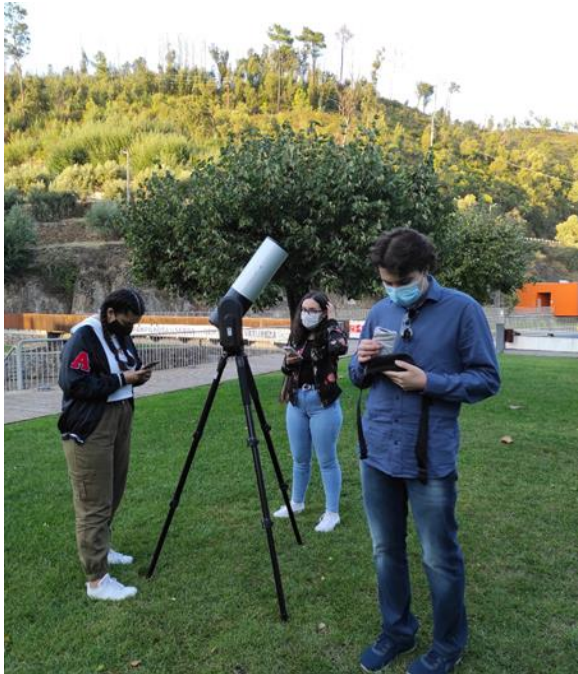

Figure 1 - Preparing for observations in the centre of Pampilhosa da Serra in COVID-19 times.

Portugal has been promoting Dark Skies related activities benefiting from its long summer with clear night skies, optimal for stargazing activities. These not only became a highly valued activity, key ingredient to the current ecotourism trend but also enabled some citizen science applications in particular related to planetary sciences like exoplanet transits, planetary defense observations as part of space situational awareness programs and observations of asteroid occultations. Thus, we are working with Pampilhosa da Serra in the Central Region in Portugal, its Intermunicipal Coimbra Region and related relevant tourism associations, to foster citizen science and astrotourism as a showcase of the Dark Sky Aldeias do Xisto at Pampilhosa da Serra territory.

The main barriers to citizen science and in particular to cut a program with many nights of observation depend on access to adequate training and basic astronomy knowledge. This requires celestial objects knowledge and identification, history of astronomy and of course the trained ability to perform observations that may be cumbersome or somewhat challenging to the neophyte with classic telescopes. These are paramount characteristics to devise an astrotour guide qualification that requires the help of professionals or acquainted amateurs and some technical means. As in other examples, [3,4], to bridge the gap towards consolidation of basics of astronomy education we developed a first citizen science program sponsored by the Portuguese Science Education program (Ciência Viva), intended for the younger students in a summer camp for about two weeks.

In the Summer of 2020, in the middle of COVID-19 pandemics, we started an initiative aiming to train young students using an Unistellar eVscope, a smart, compact and user-friendly digital telescope that offers unprecedented opportunities for deep-sky observation and citizen science campaigns. This took place as an innovative methodology for learning STEAM (Science, Technology, Engineering, Arts and Mathematics), to test digital telescopes for future astrotourim activities, its connection to social networks, and to sow the seeds for astronomy clubs in a low-income, interior region of Portugal. This activity, Cyber-Cosmos, was probably the first continuous application of this equipment in a pedagogical and citizen-science context, and in a pandemic context

2. **Choosing a digital telescope for citizen science**
The telescope used in our activities is the Unistellar eVscope (4.5-inch "enhanced Vision telescopes"). Presented at Web Summit 2019 in Lisbon, the eVscope features an integrated Wi-Fi system that connects to mobile devices within a 50-meter radius, allowing users to remotely operate the telescope and download the images through an app on their smartphones and tablets — especially useful at a time of social distancing. The idea behind our training is to foster and guide a citizen-science community, where anyone can contribute to astronomical discoveries by sharing their images in real-time with scientific observation campaigns observing planet occultations, doing planetary defense or simply stargazing. It also allows any sky watchers to enjoy simple stargazing and get beautiful images of the cosmos even in urban areas with light polluted skies.

As outlined in [7], Unistellar eVscope is a 4.5" (11.4 cm) Newtonian-like (focal length = 450 mm, magnification of 50) telescope designed specifically to work in urban and countryside environments. It is equipped with a sensor located at the prime focus of the telescope. The sensor is a CMOS low-light detector IMX224 (1/3-type, 1.27 megapixels, 12-bit, up to 60 fps) produced by Sony and characterized by a gain amplifier of up to 72 dB and a low read noise of less than 1 e- which allows us to record multiple frames with exposure times between 1 ms and 4 s. An on-board computer stacks and processes those frames (dark and background removal, shift-adding and stacking) to produce an improved image which is projected in real time through the electronics eyepiece. Each individual frame is stored in the telescope and can be accessed in 12-bit TIFF format by the user for a posteriori data processing and analysis [7]. As a major citizen science global development for astronomy and space sciences, an Unistellar eVscope Network has been formed since March 2020 with an enthusiastic worldwide network of more than 5000 citizen and professional astronomers. This worldwide spread community of citizen astronomers operates their backyard eVscopes to perform observations of transiting exoplanets and asteroid occultations under guidance of professional astronomers from the SETI Institute and in collaboration with major astronomy associations. SETI astronomers do prepare campaigns, publish observation data and operation instructions that can be followed by the citizen scientists or astro-enthusiasts simply by following their smartphone or tablet apps. A simple registration database sends automatic notifications via email or slack channel to any registered users about events visible in the sky or about requested observations by professional or amateur astronomers through the eVscope network. Namely, to describe a few examples, planetary defense activities and asteroid occultations are very important to outline encounter strategies by missions like Lucy [9], a NASA space probe launched by the 16th October 2021 that will complete a 12-year journey to several asteroids with fly-byes to a main belt asteroid and to six Jupiter trojans asteroids, which share Jupiter's orbit around the Sun. On the other hand, the high discovery throughput of exoplanet candidates from space observatories like the NASA TESS mission have created a high demand for ground-based follow up observations, that will likely increase with other space observatories like the James Webb Space Telescope, ARIEL or PLATO [11,12,13]. These space observatories will observe deeply at the most interesting exoplanets found by TESS to study their atmospheres and search for possible signs of life. The TESS space telescope alone has discovered more than 2,000 new exoplanet candidates with many requiring additional observations to confirm their planetary nature and/or refine measurements of their orbits [8]. These are activities that can be performed with smaller ground instruments and therefore can be handled by a network of citizen astronomers.

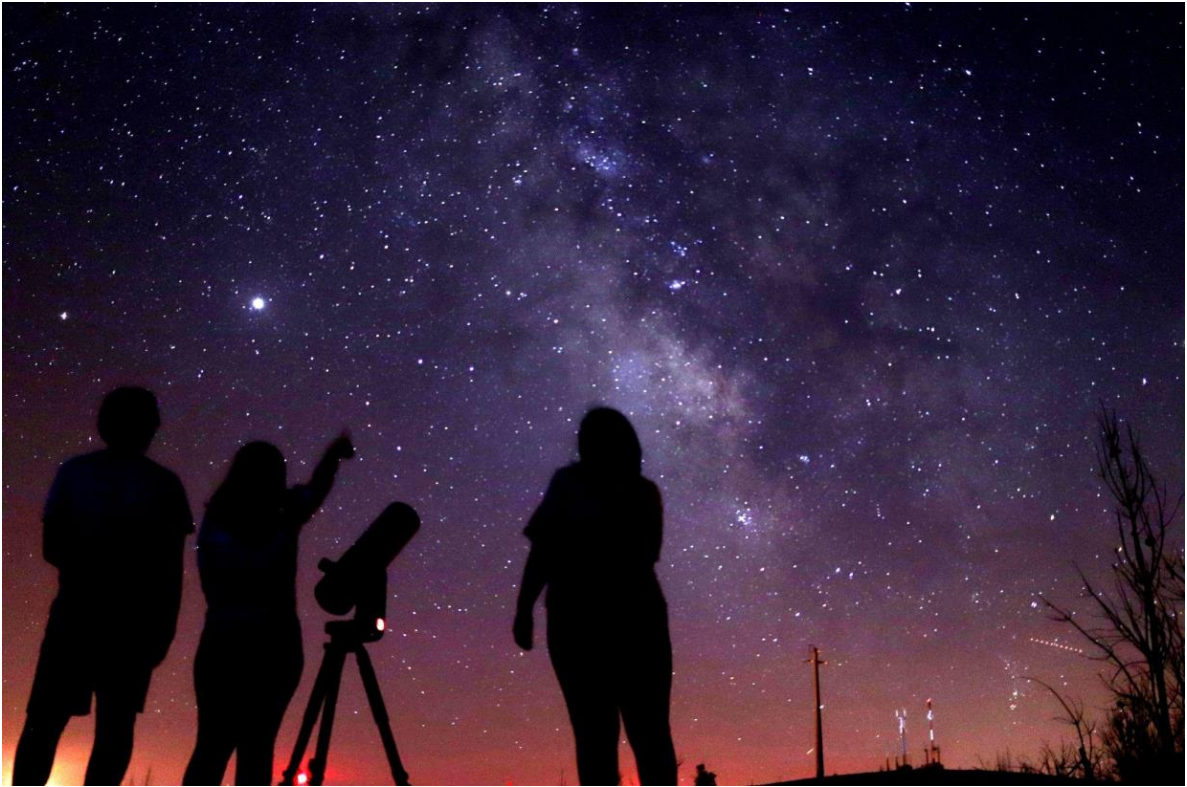

Figure 2 - Students preparing for Night sky observation with eVscope; snapshot of the Milky Way at Fajão, Pampilhosa da Serra.

3. **Observations and training of the students**

During that period of two weeks, within the Ciência Viva Summer Camp, from the 29th of July to the 12th of August of 2020, a group of three students (co-authors CM, DS and DM) engaged in a citizen science project based learning (PBL) activity [14] with about 2-3h observing sessions every night. This approach, centred on the students, focuses on teaching students through immersive and engaging projects that emphasizes autonomy rather than the more standard teacher-student relation of listen and take notes approach. Each day, during a couple of hours in the afternoon the students prepared a list of objects to observe during the night. These lists were ordered considering the elevation of the objects at a given moment, selecting brighter objects or avoiding observing objects close to the full moon. Particular attention was paid to careful observation planning, choosing the nature of the objects and the observational strategy: either planets or closer sources or deep sky observations of galaxies and nebulae requiring longer integration times. During the afternoon period the students also collected information, curiosities or interesting myths about the objects. They learnt about the objects that belong to the Solar system, the ones that are beyond but in the Galaxy, and about other galaxies. They learnt about the life cycle of stars, their different colours and sizes. More importantly, they learnt about the importance of a dark sky and its relation to the local history: night navigation was essential in a region where transhumance was in a not so distant past an important seasonal activity, and where old trade routes between mountainous villages were established using sky reference points. These lead to many stories and legends still known by the elders, a treasure that must be collected for our future cultural memories before it disappears.

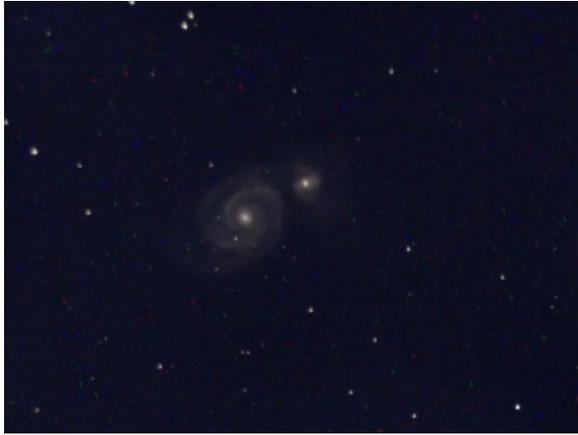

Figure 3 - Whirlpool Galaxy (M51), an example of a spiral galaxy.

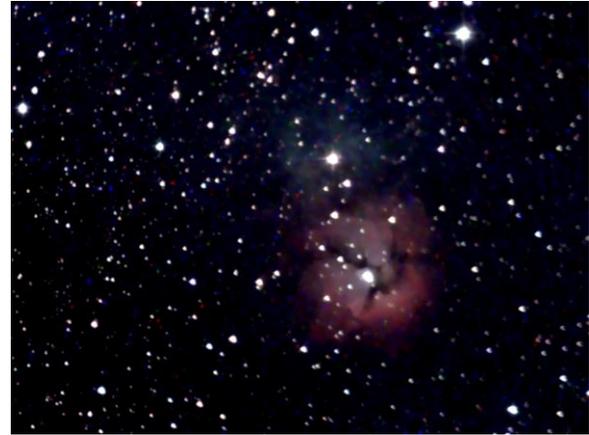

Figure 4 - Trifid Nebula (M20), an HII region in a star-forming region.

On the first night it was explained to the students how to collimate, and how to focus the eVscope. Night after night, they repeated these procedures with progressively less intervention from the supervisors. By the start of the second week they were able to operate the eVscope by themselves. Figures 2 to 7 show some examples of the deep sky images obtained by the students.

The last days/nights were dedicated to the preparation of a public observing session to be conducted by the Students and that included senior officials from the local Town Hall. This session took place on the 12th of August at Pampilhosa da Serra football stadium. The public was limited due to the pandemic restrictions, but it was demonstrated that the students were able to prepare and conduct the session by themselves, they were not only capable of operating the eVscope, but they were already able to help the people in the audience to install and use the app to get and share images in real time. It was also an opportunity to foster dialogue about light pollution and the importance of a dark sky for our cultural heritage and as an economic asset. The ease of use of the eVscope is a key factor to qualify astrotour guides in a short period of time. On the other hand, the possibility to obtain images in real time, in portable devices such as cell phones or tablets, led to excitement in the public during the last night session, but was also very useful to keep the students interested and awed during the two weeks' period. In fact, this student-centered and interactive approach with a well devised eVscope program contributes to improve student learning and knowledge acquisition, better problem solving and analytical thinking skills [15, 16].

The students are now capable of preparing public outreach observing sessions with the eVscope, but maybe even more interesting for students is the possibility to participate in the citizen science campaigns easily accessible through the eVscope app.

4. **Follow-up Citizen Science: planetary sciences**

After the exploration within Ciência Viva Summer camp of deep sky capabilities of the telescope, we have participated in a number of observational campaigns of asteroid occultations, contributing with observations to several occultation analysis and to exoplanet transit observation. Here, we participated in the campaign launched by the French Astronomy Association in June 2021 for the follow up of WASP-148b exoplanet transit. We have followed a simple set of instructions that were easily understood by the occasional astronomer. Although the data was not enough to cover the full transit since clouds affected data acquisition in the second half of the transit, the data retrieved already shows the potential of citizen science with this digital telescope with automated FOV recognition (Figure 9).

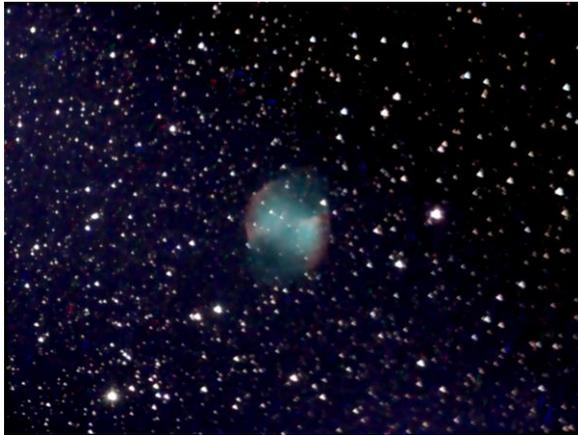

Figure 5 - Dumbbell Nebula (M27), a planetary nebula.

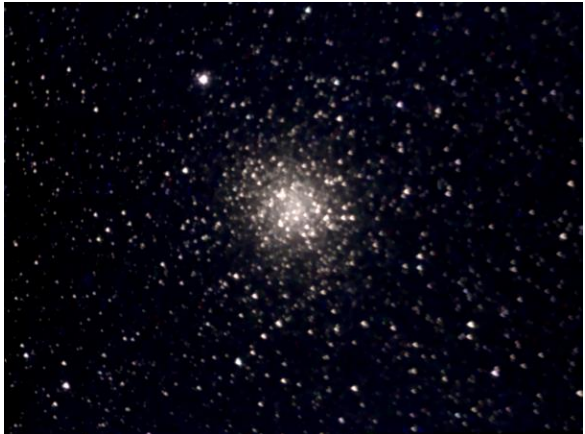

Figure 6 - Globular cluster (M22).

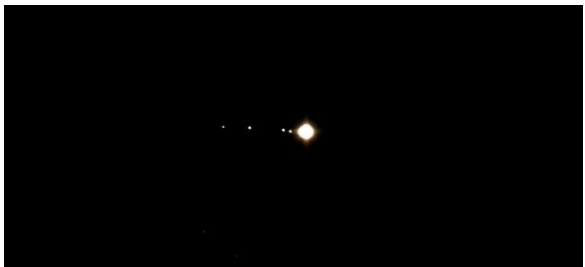

Figure 7 - Jupiter and the Galilean satellites.

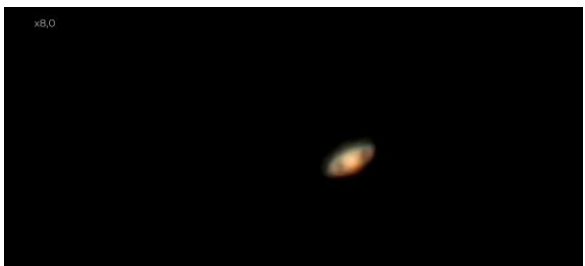

Figure 8 - Saturn.

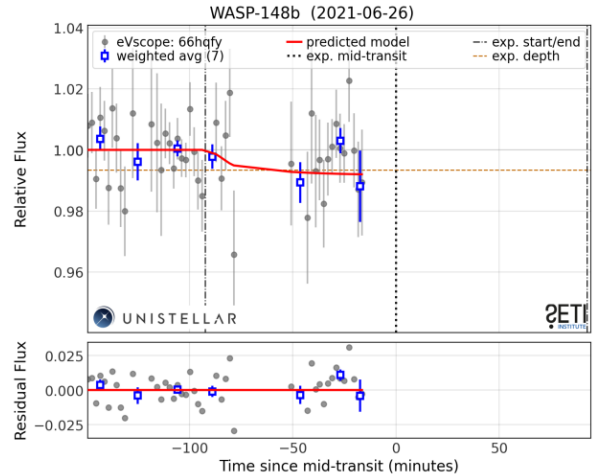

Figure 9 - Exoplanet transit observation of WASP-148b at Pampilhosa da Serra Space Observatory (PASO): Campaigns led by the French Astronomy Association, with Unistellar eVscope network sponsorship. Observation carried by the team. Only half of the target was observed since clouds affected data acquisition in the second half of transit observations. (Tom Esposito, SETI, June 2021).

## 5. Conclusions and Future Work

Astrotourism and related citizen science activities are central to protection of the skies and space science education. In the Summer of 2020, in the middle of COVID-19 pandemics, we started an initiative to train young students using an Unistellar eVscope, a smart, compact and user-friendly digital telescope that offers unprecedented opportunities for deep-sky observation and citizen science campaigns. This was probably the first continuous application of this equipment in a pedagogical and citizen-science context, and in a pandemic context.

We remark that these activities can be handled by a network of citizen astronomers taking advantage of dark nights with low light pollution like the Dark Sky Aldeias do Xisto region in central Portugal. In doing so, they contribute themselves to raise awareness of light pollution and foster a dialogue with local towns to dim their night lights without prejudice to their activities. The digital, automatic FOV recognition enables anyone, even with a modest knowledge of the sky, to contribute to citizen science initiatives with fast learning pace. With the establishment of PASO dedicated to Space Surveillance Awareness activities, we expect a steady support to observation campaigns of asteroid occultations, planetary defense, exoplanet transits and to the promotion of stargazing parties as part of local ecotourism activities. On the other hand, this PBL, person-centred and interactive approach has produced great enthusiasm and led to excellent learning autonomy skills. The simplicity of this telescope and observation planning enriches STEAM activities and may greatly benefit future follow-up classroom-type activities within Modelling Instruction scenarios (inquiry-based science pedagogy).

**Acknowledgements**

CM, DS and DM acknowledge financial support from Ciência Viva. We warmly thank the City Hall of Pampilhosa da Serra for all the support and logistics. We warmly thank ATLAR Innovation and Unistellar for support and further discussions. The team acknowledges further support from ENGAGE-SKA Research Infrastructure, ref. POCI-01-0145-FEDER-022217 and PHOBOS (POCI-01–0145-FEDER-029932), funded by COMPETE 2020 and FCT, Portugal; exploratory project of reference IF/00498/2015 funded by FCT, Portugal. IT team members acknowledge support from Projecto Lab. Associado UID/EEA/50008/2019. A.C. acknowledges support from CFisUC projects UIDB/04564/2020 and UIDP/04564/2020. JP and BC have been supported by the European Commission H2020 Programme under the grant agreement 2-3SST2018-20.